\documentclass[apj]{emulateapj}

\usepackage{graphics}
\usepackage{epsfig}

\shorttitle{High velocity HI is not associated with W51C}
\shortauthors{Tian \& Leahy}

\begin{document}

\title{High velocity HI is not associated with TeV Supernova Remnant W51C}

\author{W.W. Tian\altaffilmark{1,2}, D.A. Leahy\altaffilmark{2}}
\altaffiltext{1}{National Astronomical Observatories, CAS, Beijing 100012, China. tww@bao.ac.cn}
\altaffiltext{2}{Department of Physics \& Astronomy, University of Calgary, Calgary, Alberta T2N 1N4, Canada}
 
\begin{abstract}
The recently-detected TeV $\gamma$-ray source HESS J1923+141 
coincides with  
Supernova Remnant (SNR) W51C and the star forming region W51B of 
the W51 complex. We construct HI absorption spectra to SNR W51C, 
HII regions G49.2-0.35 and G49.1-0.38 in W51B, and a nearby compact 
extragalactic source.  
Our study detects high-velocity (HV) HI clouds (above 83 km 
s$^{-1}$) which coincide with W51B, but finds that the clouds are
behind W51B. Both W51C and G49.2-0.35 have have similar highest-velocity
absorption features at $\sim$70 km s$^{-1}$. 
The HII region G49.1-0.38 is behind the SNR because its HI 
absorption spectrum has a feature at 83 km s$^{-1}$. 
These new results argue against previous claims that the SNR 
has shocked the HV HI clouds. Therefore the TeV  
emission from the complex should not be associated with the HV HI 
clouds. W51C has a distance of about 4.3 kpc, smaller than the 
tangent point distance of 5.5 kpc in that direction, but still in 
the Sagittarius spiral arm.    
\end{abstract}


\section{Introduction and Data}

It is widely known that most Galactic cosmic Rays(CRs) probably 
originate from supernova remnants (SNRs). 
Recent studies have supported that young SNRs ($\sim$ 1000 yr old) 
accelerate charged particles (protons and electrons) in the 
interstellar medium (ISM) and circum-stellar medium (CSM) 
up to 10$^{15}$ eV (e.g. Bell et al. 2013). However, it is still 
unclear whether the charged particles emitting the very-high-energy
$\gamma$-rays are accelerated protons or electrons, i.e. whether the 
$\gamma$-rays are of hadronic origin or of leptonic origin. 
Recent progress in high energy and Very-High-Energy (VHE) 
$\gamma$-ray observations (Abdo et al. 2010; Helder et al. 2012) and 
in theoretical investigations (Ohira, Murase and Yamazaki 2011; Li \& 
Chen 2010, Malkov et al. 2011) suggests that the resolution of the
hadronic/leptonic ambiguity will soon occur. A key part of this 
resolution is observation and interpretation of GeV-TeV-bright 
middle-aged SNRs interacting with molecular clouds. 

W51C is one of several such interacting SNRs (W28, W44, W51C and 
IC443) which have recently been studied in the context of particle acceleration mechanisms (Li \& Chen 2012, 
Aleksic et al. 2012, Fang \& Zhang 2010, Abdo et al. 2009).
W51C is located in the complex W51, one of the strongest radio 
sources in the Galaxy. The W51 complex contains components of thermal emission
from HII regions (W51A and W51B) and non-thermal emission from SNR W51C 
(Kassim 1992, Seward 1990). 
W51C is a strong X-ray emitter 
(Koo et al. 1995, 2002, 2005; Hanabata et al. 2012). It has been 
evident that the soft X-rays are thermal and most 
likely from the SNR interior, and part of the hard X-rays are from a 
compact object within the SNR, and another part related with stellar winds in the HII regions. 
Recent $\gamma$-ray observations have revealed that the complex is a 
strong GeV-TeV source (Aleksic et al. 2012), with emission 
coincident with W51B and W51C but not W51A. It is unknown whether the 
VHE emission is associated with the SNR W51C or originates from the star forming region W51B.

In this letter, we use HI absorption lines to study the distances 
of W51C and the HII regions of W51B and to determine whether high 
velocity HI is associated with any of these objects.
We discuss possible origin of TeV $\gamma$-ray emission in the context of the distance results.
We utilize the 21 cm continuum and HI line data from the VLA Galactic Plane Survey (VGPS, see details in Stil et al. 2006) in the direction of the W51 complex.  

\section{Analysis and Results}

\subsection{Continuum images and Spectra}
The methods to extract HI absorption spectrum are detailed in  Tian, Leahy \& Wang (2007) and  Leahy \& Tian (2010). 
In general, the HI absorption spectrum of a radio source can be found by the formula: 
$\Delta T$ = $T^{HI}_{off}$-$T^{HI}_{on}$ = ($T^{c}_{s}$-$T^{c}_{bg}$)(1-$e^{-\tau}$). 
$T^{HI}_{on}$ and $T^{HI}_{off}$ are the average brightness temperature of many spectra from 
a selected area on a strong continuum emission region of the target source and an adjacent 
background region (i.e. excluding the strong continuum emission area).  
$T^{c}_{s}$ and $T^{c}_{bg}$ are the average continuum brightness temperatures for the 
same regions respectively. $\tau$ is the optical depth from the continuum source to the 
observer along the line-of-sight.  The main uncertainties in the absorption spectrum 
are caused by the differences in distribution of small HI clouds along the line-of-sight to 
source and background regions. Our methods are able to 
minimize the differences in HI distribution between the source 
region and the background regions by choosing them to be adjacent.

We show the 1420 MHz continuum image of the W51 complex in Fig. 1. 
Fig. 1 also shows 3 large ellipses which roughly indicate the areas of W51A W51B and W51C, 
and 4 small circles where HI-line spectra are extracted. 
The small circles cover the HII regions G49.2-0.35 and G49.1-0.38 (bright HII regions in W51B),
a bright region in W51C, and a nearby compact source (CS) G49.21-0.96. 

The HI spectra of the bright HII regions G49.2-0.35 (T$_{B}$ $\sim$ 800 K) and G49.1-0.38 
(T$_{B}$ $\sim$ 255 K) are shown in the top panels of Fig. 2. 
The highest velocity  absorption peak is at 65$\pm$8 km s$^{-1}$ for G49.2-0.35 and 
at 75$\pm$8 km s$^{-1}$ for G49.1-0.38. The shapes of the two HI absorption spectra are similar 
except for the absorption peak at 75$\pm$8 km s$^{-1}$, which only appears for G49.1-0.38. 
Some low T$_{B}$ HI clouds ($\sim$10 K) with forbidden velocities of 80 $-$ 120 km s$^{-1}$ 
appear in G49.1-0.38's HI emission spectrum and those at velocity of above 83 km s$^{-1}$ 
have no associated absorption. 
We calculate absorption column densities from the HI absorption spectra: 
N$_{HI}$ $\sim$ 1.21 $\times$ 10$^{22}$ cm$^{-2}$ for G49.2-0.35, 
and $\sim$1.27 $\times$ 10$^{22}$ cm$^{-2}$ for G49.1-0.38. 
We take a typical value of T$_{s}$= 100 K, N$_{HI}$= 1.9 $\times$ 10$^{18} \tau \Delta{\it{v}} T_{s}$ cm$^{-2}$, Dickey \& Lockman 1990).     

The W51C spectrum (3rd panel of Fig.2) is from a faint source so there is much more noise in its absorption spectrum. 
The main absorption features at 5, 25, 47 and 65 km s$^{-1}$ in the HII regions' absorption spectra 
are all clearly seen in the W51C spectrum. 
The highest reliable absorption velocity towards W51C is 70$\pm$3 km s$^{-1}$, close to that of
 G49.2-0.35 and lower than that of G49.1-0.38. 
 The absorption column density of W51C is $\sim$0.45$\times$10$^{22}$cm$^{-2}$, smaller than 
 that of either G49.2-0.35 and G49.1-0.38. 
 The above two facts show that W51C is likely near to G49.2-0.35 but is in front of G49.1-0.38 
 (i.e. part of HI clouds in velocity range of 73 - 83 km s$^{-1}$ are between W51C and G49.1-0.38). 

From the HI spectrum of CS G49.21-0.96 (T$_{B}$ $\sim$170 K, 4th panel of Fig.2), we see 
two absorption features at velocities of -42 and -53 km s$^{-1}$, which 
show the source is an extra-galactic source. In addition, the highest velocities of emission 
(T$_{B}$ $\sim$20 K) and absorption towards the compact source are both at 79$\pm$4 km s$^{-1}$. 
Table 1 summarizes the HI maximum-velocity absorption/emission features of the four sources.

\begin{figure} 
\vspace{50mm} 
\begin{picture}(80,80)
\put(-55,230){\includegraphics{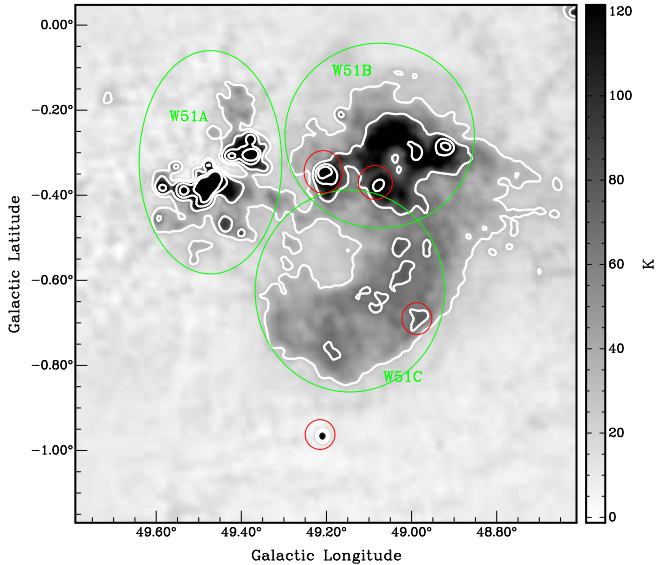}}
\end{picture}
\caption{The 1420 MHz continuum image of the W51 complex, with contours at 30, 70, 180 and 380 K. W51 A, B and C are indicated by the three large ellipses (green in the online version).  The small circles (red in the online version) show the 4 regions from which we extract HI spectra.}  
\end{figure} 

\begin{figure*}
\vspace{195mm}
\begin{picture}(80,80)
\put(0,485){\includegraphics{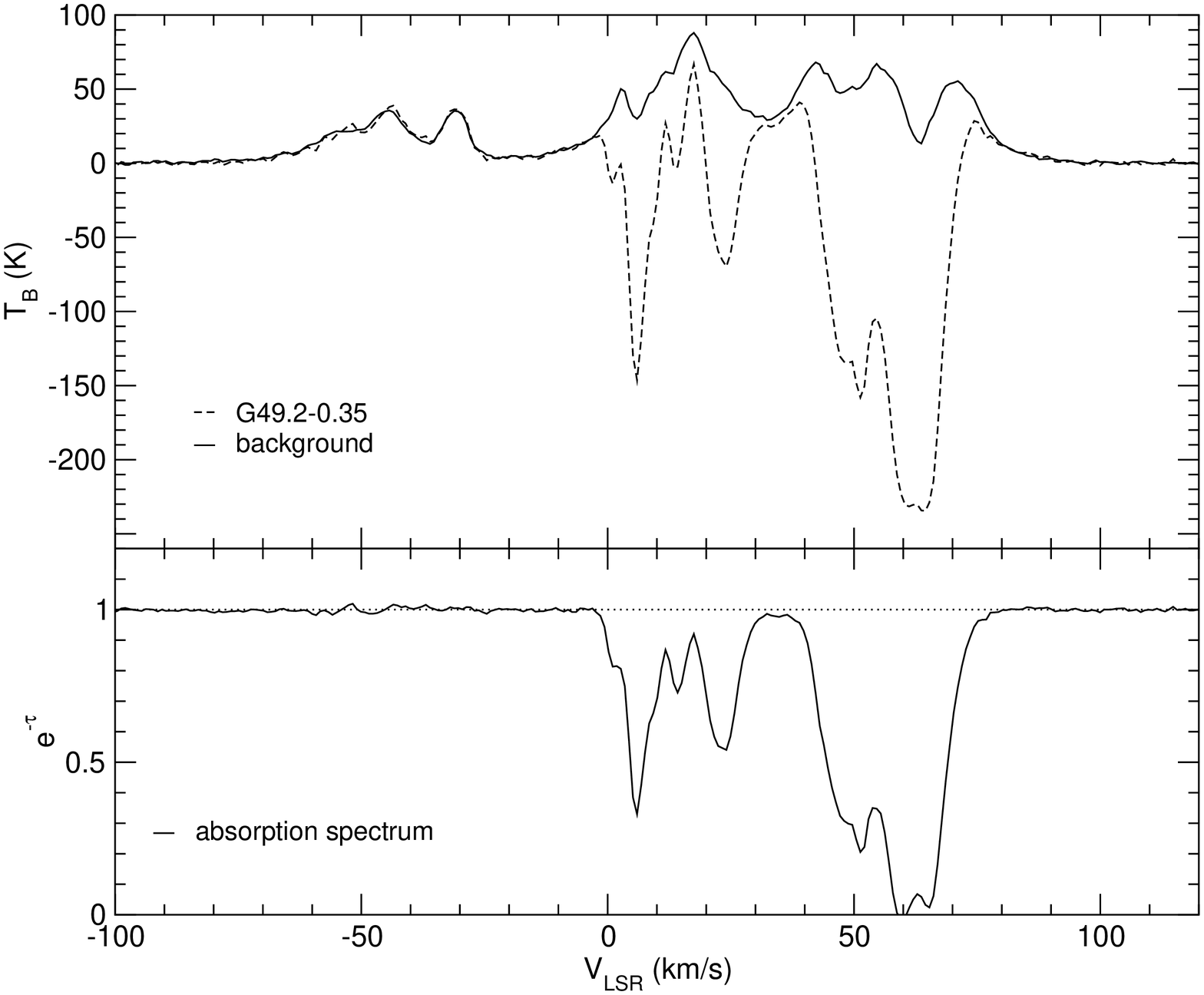}}
\put(0,325){\includegraphics{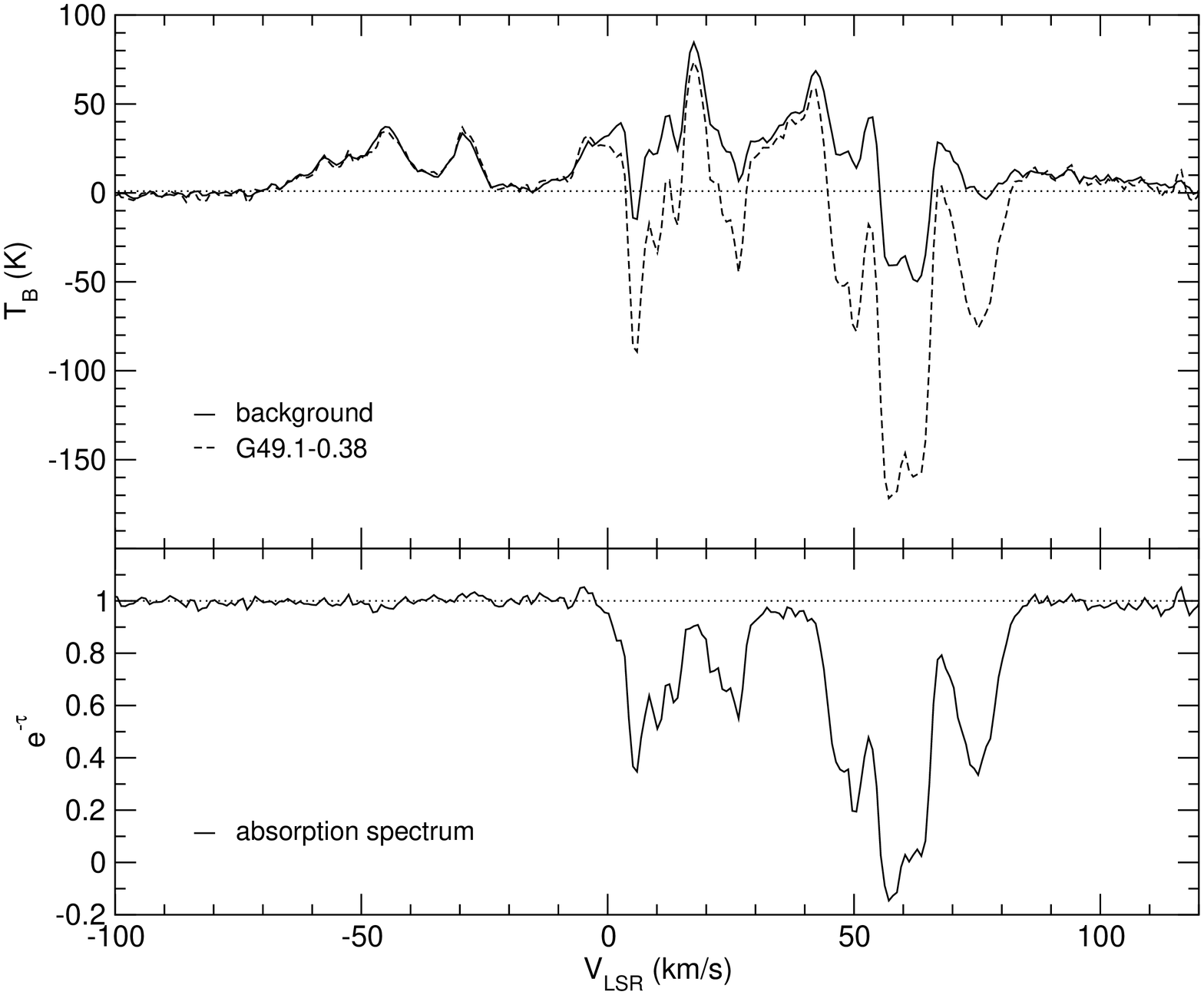}}
\put(0,165){\includegraphics{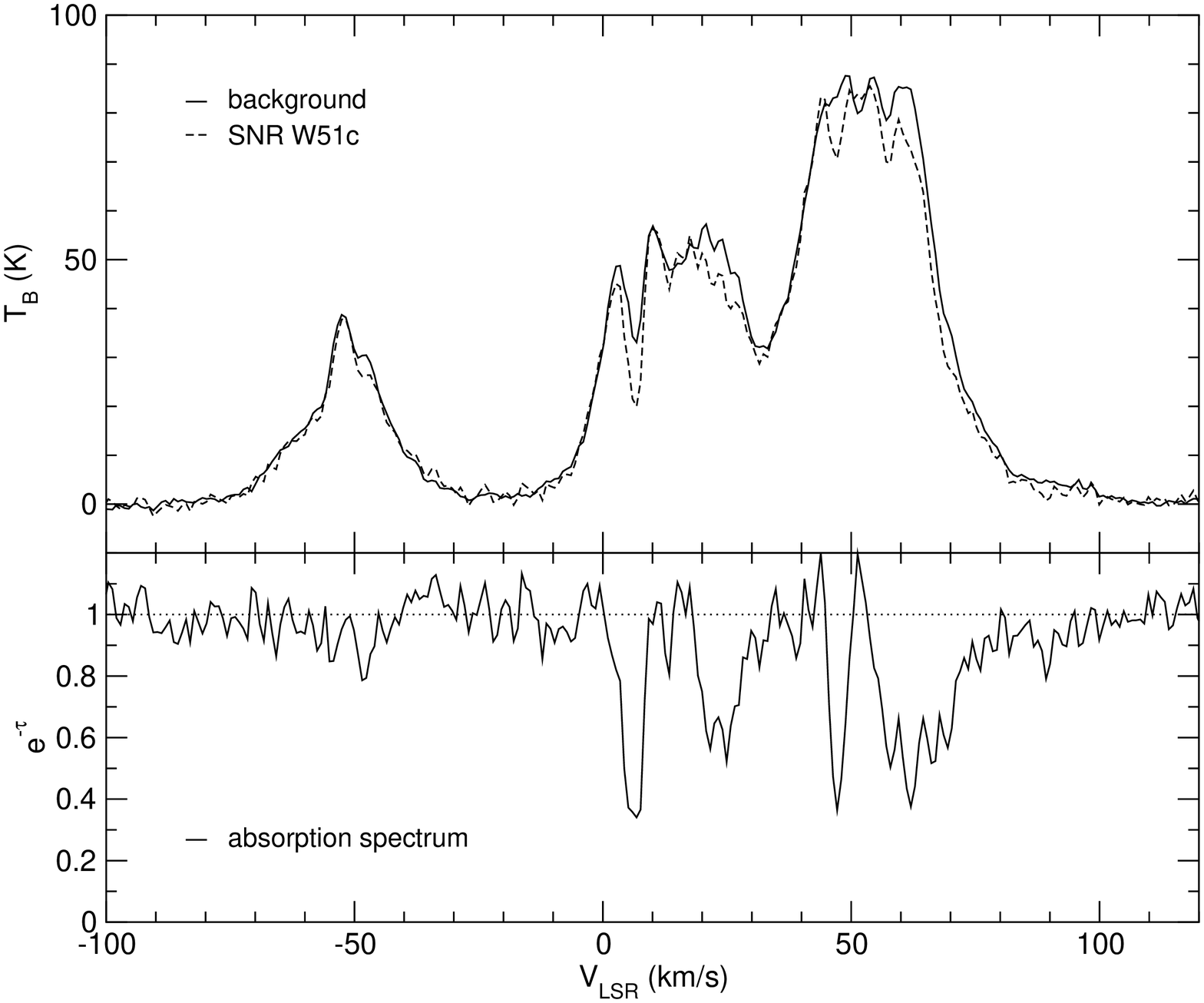}}
\put(0,5){\includegraphics{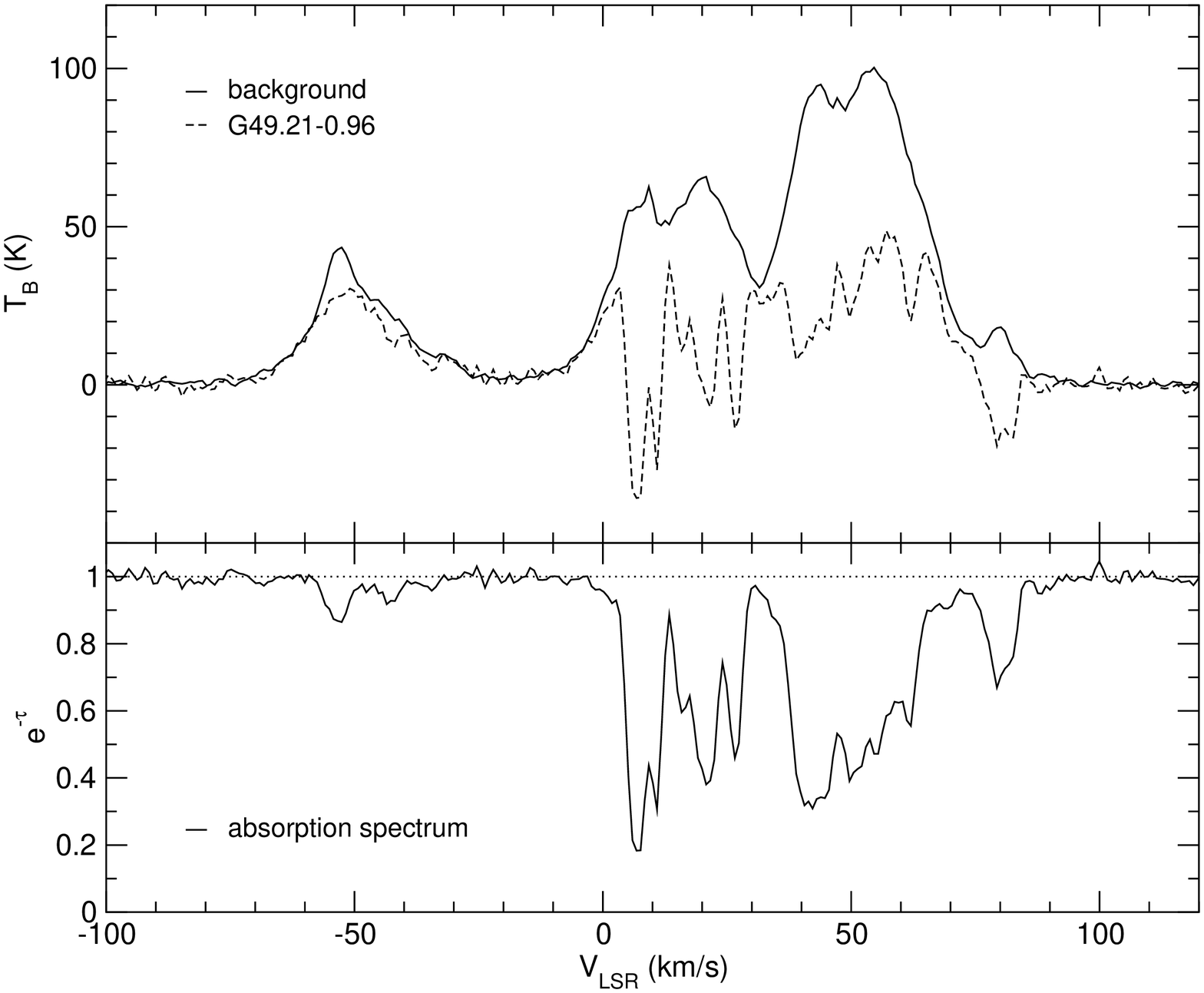}}
\end{picture}
\caption{HI emission (upper half of each panel) and respective HI absorption (lower half of each panel)
 of the two HII regions G49.2-0.35 and G49.1-0.38, of SNR W51C and of the extragalactic compact source G49.21-0.96.}
\end{figure*}

\begin{table}
\begin{center}
\caption{HI absorption/emission features of four sources}
\setlength{\tabcolsep}{1mm}
\begin{tabular}{ccccc}
\hline
Source Name: &  G49.2-0.35 & G49.1-0.38 & W51C & G49.21-0.96 \\
 & HII & HII & SNR & nearby CS \\
\hline
V$_{MA}$(km s$^{-1}$) & 65$\pm$8 & 75$\pm$8 & 70$\pm$3 & 79$\pm$4 \\
V$_{ME}$(km s$^{-1}$) & 72$\pm$8 & $\ge$115 & 70$\pm$8 & 79$\pm$4 \\
\hline
\end{tabular}
\end{center}
V$_{MA}$:maximum absorption velocity, V$_{ME}$:maximum emission velocity.\\
\end{table}

\subsection{HI Channel maps}

\begin{figure}
\vspace{190mm}
\begin{picture}(80,80)
\put(-40,620){\includegraphics{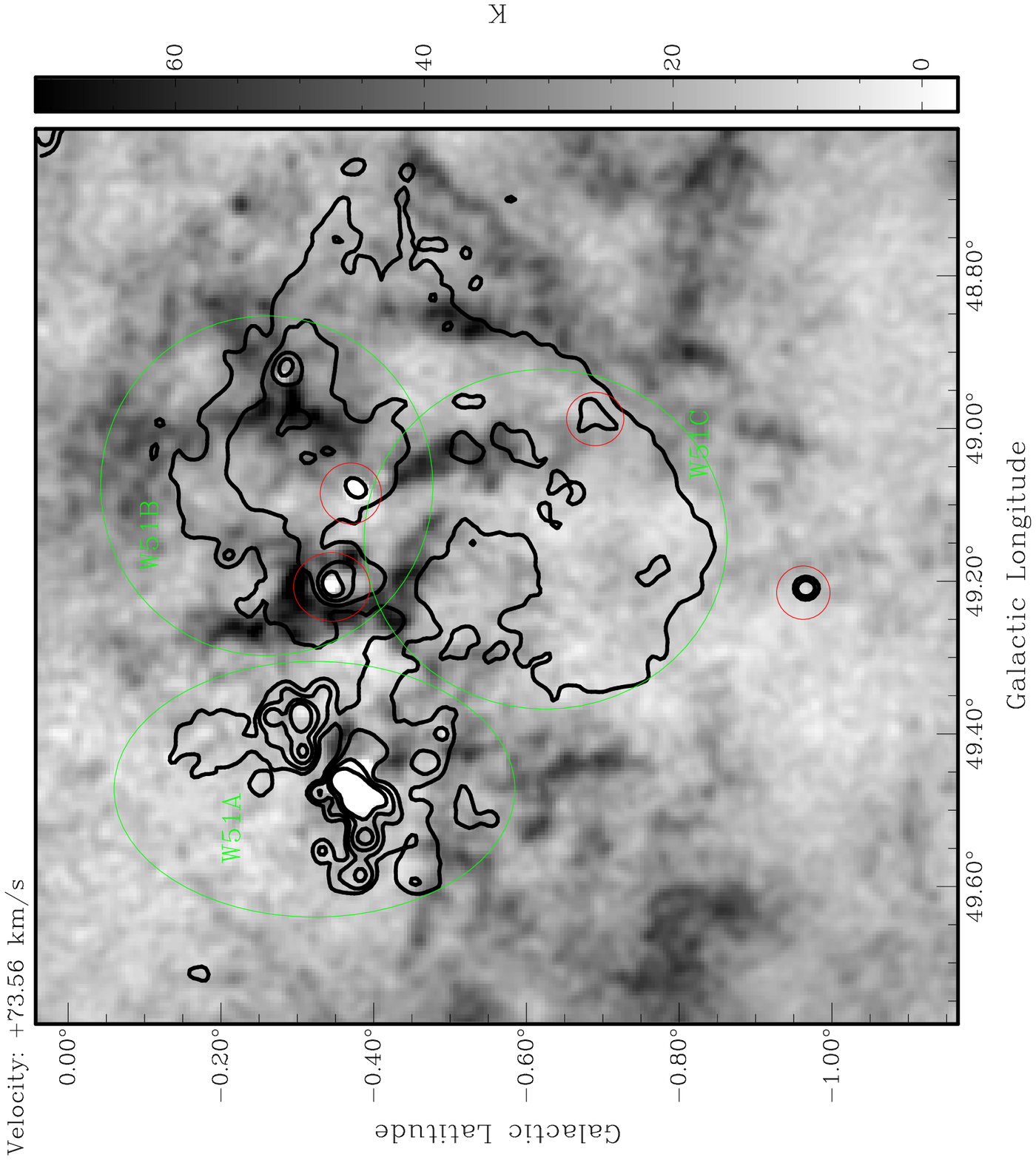}}
\put(-40,420){\includegraphics{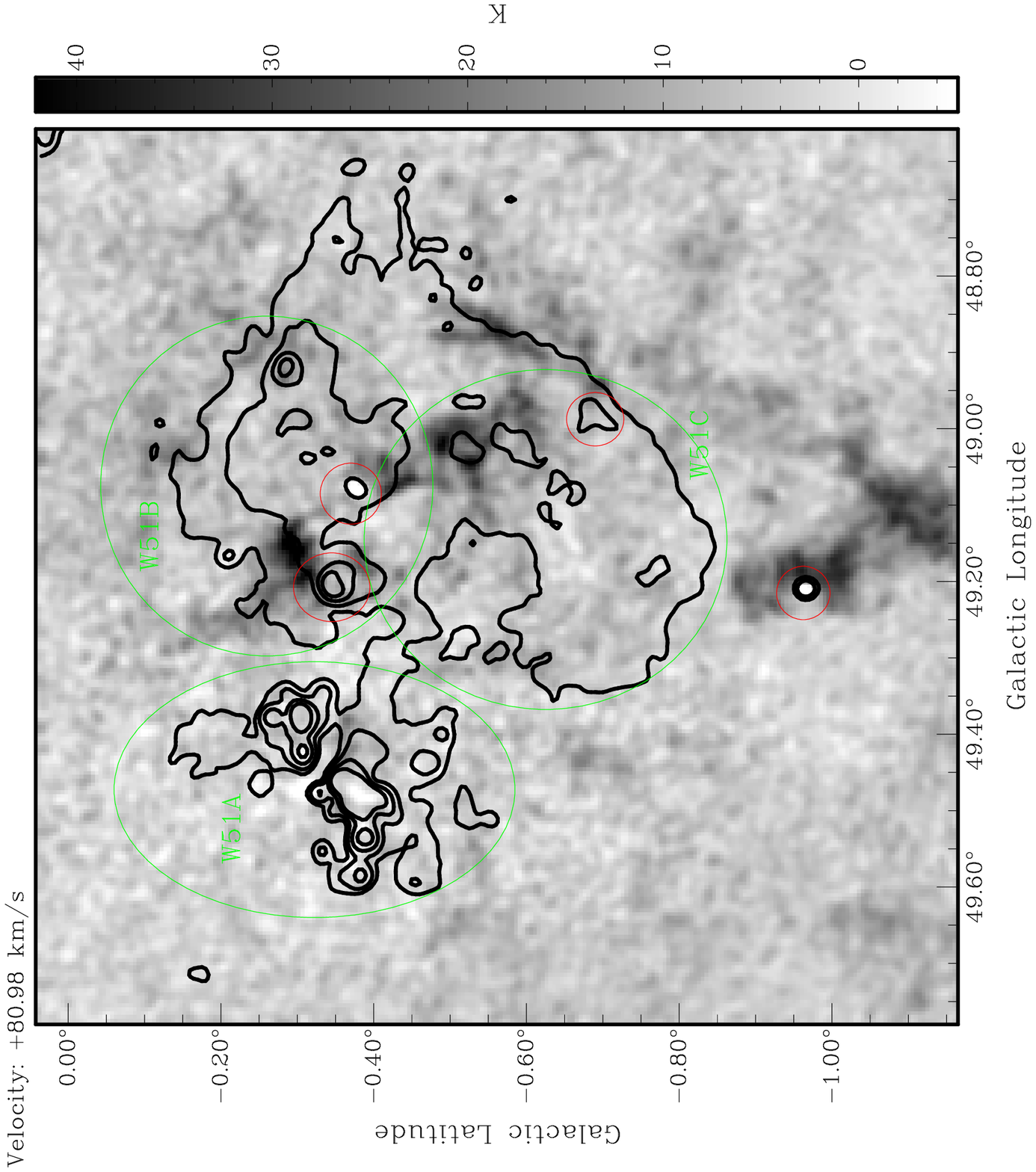}}
\put(-40,220){\includegraphics{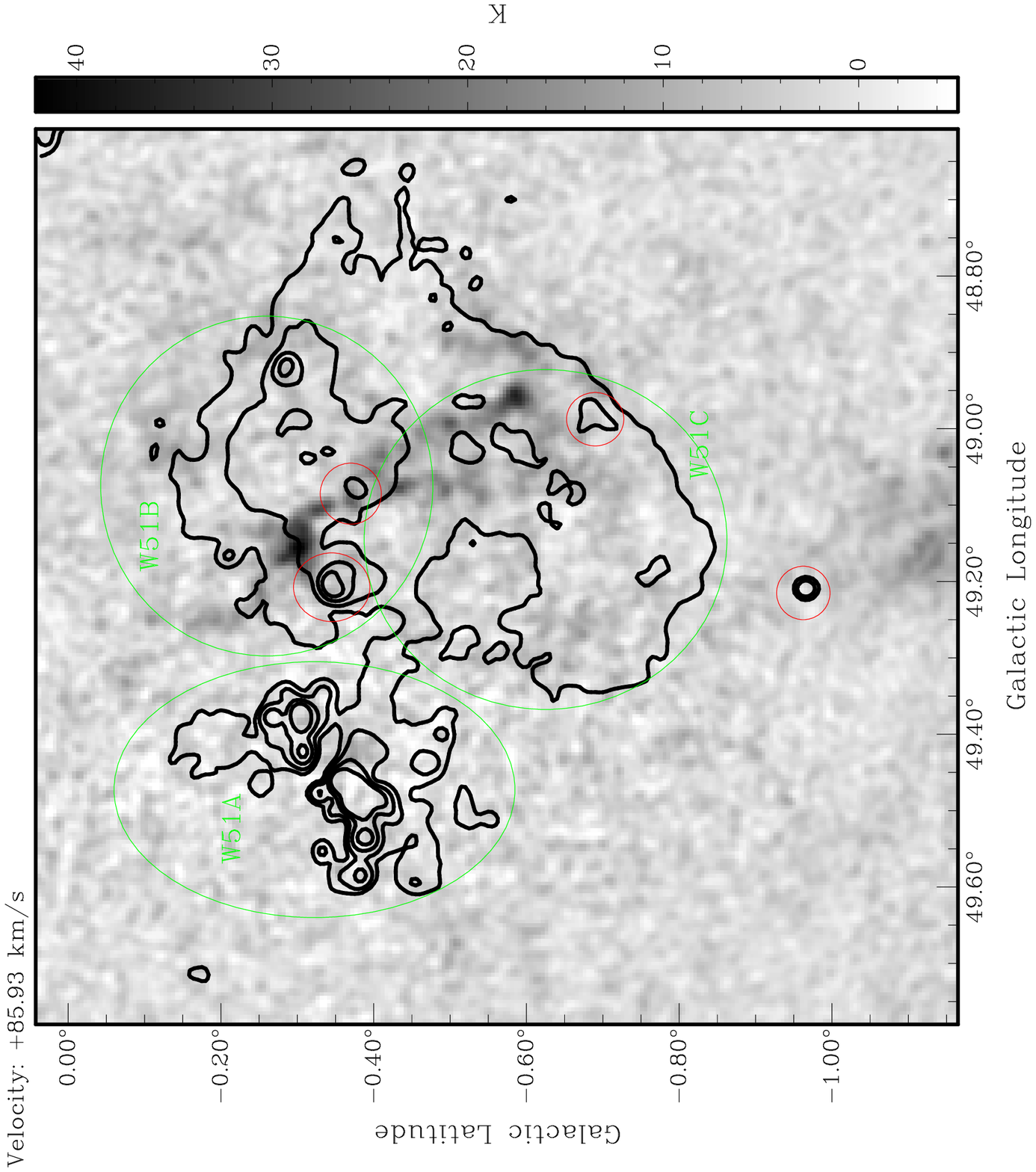}}
\end{picture}
\caption{The panels show the HI channel maps at velocities of 73, 81 and 86km s$^{-1}$ (velocity label at top left of each panel). 
Overlaid in all panels is the 1420 MHz continuum emission with contours at 30, 70, 180, 380 K. 
The HI map above 83 km s$^{-1}$ shows high-velocity extended HI clouds overlapping W51B (see text for detail).} 
\end{figure}

\begin{figure} 
\vspace{50mm} 
\begin{picture}(80,80)
\put(0,-10){\includegraphics{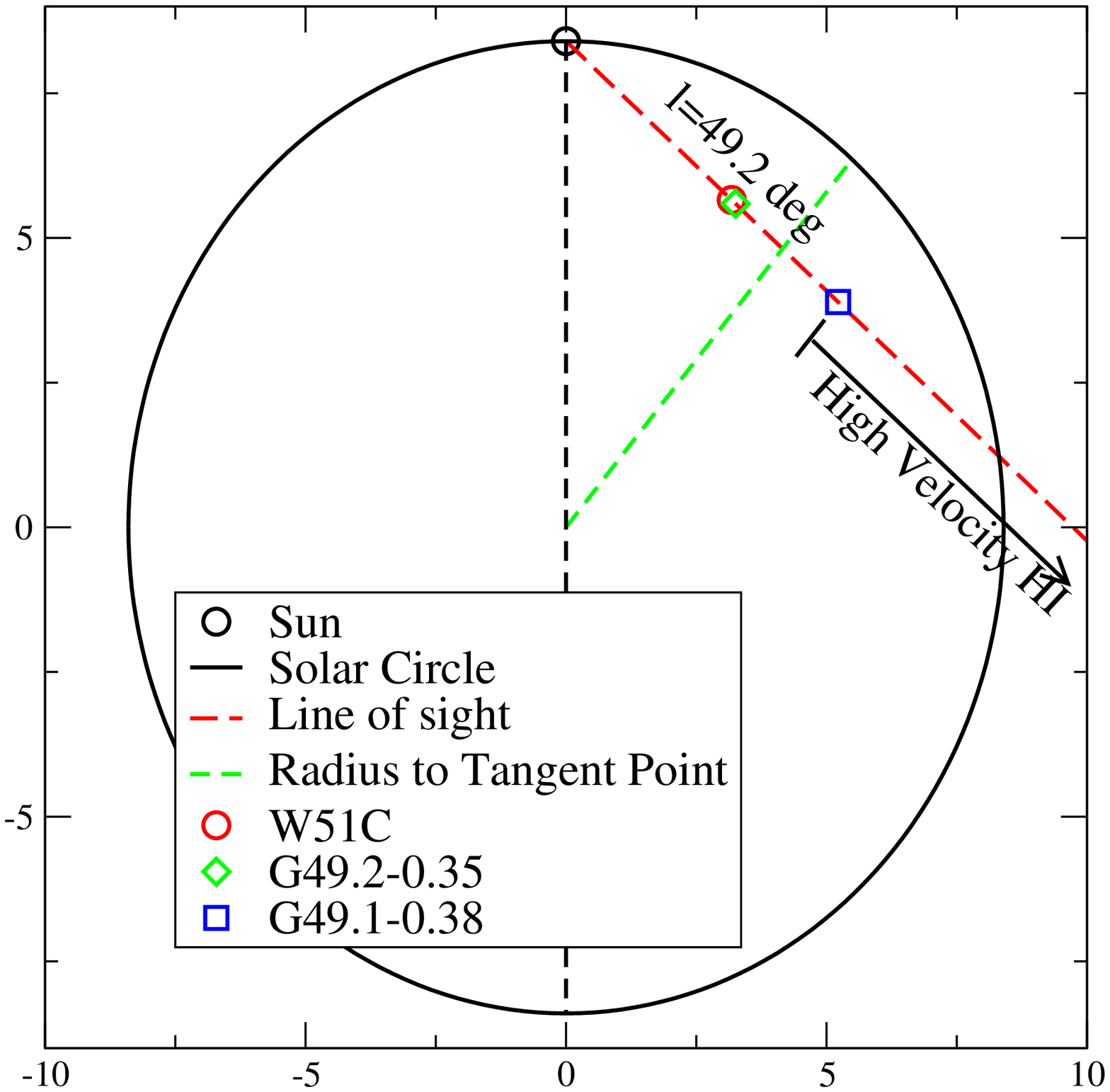}}
\end{picture}
\caption{Diagram indicating the locations of SNR W51C and HII regions  G49.1-0.38 and G49.2-0.35.
The possible range of distance for the (unknown) position of the high velocity HI is indicated by the arrow.}  
\end{figure} 

The above results are confirmed by the HI emission channel maps. In Fig. 3, we show three
selected HI channel maps towards the W51 complex. 
HI absorption towards G49.1-0.38 appears (seen as the low T$_{B}$ white spot) in the channel maps 
for velocities 80.98 km s$^{-1}$ or less, but not for velocities 85.93 km s$^{-1}$ or higher. 
The HI absorption towards G49.2-0.35 appears in the 73.56 km s$^{-1}$ (and lower) channel maps 
but not in the higher velocity maps. 
The channel map at 80.98 km s$^{-1}$ shows several high-T$_{B}$ HI clouds extending across W51B, W51C and G49.21-0.96 which show no spatial coincidence with the W51C SNR shell. 
We conclude this gas is not related with the SNR shock. 
The channel maps at velocities above 83 km s$^{-1}$ (like the map shown for 86 km s$^{-1}$)
reveal that the high velocity (HV) HI clouds overlap W51B but do not produce any associated HI 
absorption towards either G49.1-0.38 or G49.2-0.35. 
This indicates that the HV HI gas is further than either G49.1-0.38 or G49.2-0.35.

\section{Discussion and Conclusion} 

\subsection{Distances of W51C and HII regions G49.2-0.35 and G49.1-0.38}
The most distant reliable absorption features observed against the 
complex W51B and W51C are at 83 km s$^{-1}$ for G49.1-0.38 and 73 km 
s$^{-1}$ for both W51C and G49.2-0.35. They are much greater than 
the maximum velocity (i.e. tangent point velocity) of V$_{T}$
$\sim$62 km s$^{-1}$ permitted by the flat Galactic circular 
rotation curve model (taking the recently-measured parameters of 
V$_{0}$= 254$\pm$16 km s$^{-1}$, R$_{0}$ = 8.4$\pm$0.6 kpc by VLBI 
observations, Reid et al. 2009) in the direction of $\l$= 49.2$^{o}$.
However, the additional velocity of $\sim$ 10 km s$^{-1}$ could be 
due to peculiar motions (e.g. spiral arm velocity perturbations; 
streaming and random gas motions). 
In this case, we estimate V$_{T}$ $\sim$75 km s$^{-1}$ by comparing with the 
HI background emission spectrum towards $\l$=49.2$^{0}$ (which is also consistent with 
the high velocity-resolution Galactic Arecibo L-band Feed Array 
survey data, GALFA\footnote{http://www.naic.edu/$^\sim$igalfa}). 
Then the absorption features in the velocity range of 75 to 83 
km s$^{-1}$ (see HI absorption spectra of G49.1-0.38 and 
G49.21-0.96) are likely produced by the HI gas in the 
Sagittarius spiral arm with additional peculiar motions of 8 km 
s$^{-1}$ (Xu et al. 2009, Shaver et al. 1982).

It is difficult to determine accurate distances to Galactic objects 
without better knowledge of the rotation curve and non-circular 
motions. For a simplicity, we estimate the kinematic distances to 
this complex using a flat rotation curve: i.e 254 km s$^{-1}$, 
R$_{0}$=8.4 kpc, also considering V$_{R}$ linearly increasing to 267 
km s$^{-1}$ as R decreases to its value at the tangent point 
(R$_{T}$ = 6.4 kpc). Our results are: G49.1-0.38 is located 
at the far side of its Radio Recombination Lines (RRL) velocity
(V$_{RRL} = $ 67.9 km s$^{-1}$, Lockman 1989), i.e. at 6.9 kpc, 
because of the HI absorption feature at the tangent point. 
G49.2-0.35's highest absorption velocity of 65$\pm$8 km s$^{-1}$ is consistent 
with its V$_{RRL}$(67.2$\pm$1 km s$^{-1}$, Lockman 1989) 
and is smaller than the tangent point velocity. Also two OH masers 
associated with G49.2-0.35 have velocities of 69.2 and 72.1 km 
s$^{-1}$ (Green et al. 1997, Hewitt et al. 2008). 
Thus G49.2-0.35 is at the near side distance for the averaged maser velocity 
of 70.7 km s$^{-1}$, i.e., at 4.3 kpc. 
W51C's highest absorption velocity is $\simeq$ 70 km s$^{-1}$ thus it is likely close to 
G49.2-0.35, so its distance is approximately at 4.3 kpc too.
But W51C should be in front of G49.2-0.35 because of its lower column density.
The line-of-sight at $l$=49.2$^{0}$ goes through the Sagittarius arm and has about
3 kpc length inside the arm (see Figs. 9 and 10 of Hou et al. 2009). 
The tangent point distance at $l$=49.2$^{0}$ is at 5.5 kpc (also 
see Sato et al. 2010), so both W51C and G49.2-0.35 should be located at 
near side of the 3 kpc path while G49.1-0.38 is at far side. This is 
consistent with previous claims that the W51 complex, as an active 
star forming region, is located in the Sagittarius spiral arm.

The above leads to the following geometrical relation: SNR W51C and the HII region G49.2-0.35 
are at 4.3 kpc and the HII region G49.1-0.38 is at 6.9 kpc, well behind them but still 
within the solar circle. Fig. 4 here illustrates the arrangement of W51C and the HII regions
along the line-of-sight at $l$=49.2$^{0}$.

An HI 21-cm absorption line study on the W51 complex has previously
been done employing VLA and Arecibo 305m data (Koo 1997, Koo \& Moon
1997). Koo (1997) obtained HI absorption spectra of both G49.2-0.35
and G49.1-0.38 (see his Fig. 3 with source names G49.2-0.3 and
G49.1-0.4), which are roughly consistent with our spectra (Fig. 2)
except the data quality is poorer (lower sensitivity and velocity
resolution). We detect both the major HI absorption peak at $\simeq$
65 km s$^{-1}$ for G49.2-0.35, and the highest velocity absorption
peak of $\simeq$ 75 km s$^{-1}$ for G49.1-0.38. The systemic velocity of HII
G49.2-0.35 revealed by the RRL (67.2 km s$^{-1}$) and recent
CO observations (71.2 km s$^{-1}$, clump A of Ceccarelli et al. 2011)
is associated with the HI absorption feature at
65$\pm$8 km s$^{-1}$. The RRL velocity of 67.9 km s$^{-1}$
towards G49.1-0.38 is smaller than its highest HI absorption
velocity of $\simeq$ 75 km s$^{-1}$. This can be simply explained
because G49.1-0.38 is at the kinematic far side for its velocity of
67.9 km s$^{-1}$. This is consistent with our above conclusion that
G49.1-0.38 is behind of G49.2-0.35, because the HI gas between them
produces the HI absorption feature at 75$\pm$8 km s$^{-1}$ in the
G49.1-0.38 absorption spectrum.

\subsection{Is the W51C SN shock interacting with any HI or molecular clouds?}   

Previous work has shown HI absorption from HI clouds, 
with similar T$_{B}$ to the HV HI here, is clearly detectable 
(Fig.2d of Tian \& Leahy 2012, Fig. 1d of Tian \& Leahy 2008).
The HV HI patch is not detected in the absorption spectrum of G49.1-0.38
(Fig. 2, second panel), implying that it is behind G49.1-0.38.

Koo and Moon (1997) claimed detection of high-velocity HI gas 
with velocities of 85 $\sim$ 180 km s$^{-1}$ towards W51B (including 
G49.2-0.35 and G49.1-0.38, see Fig. 4 of their paper). 
We confirm detection of emission from the HV (with velocity 83 km s$^{-1}$ and higher) HI 
patch toward G49.1-0.38 (Fig. 2, second panel) but it is not seen toward G49.2-0.35 
(Fig. 2, first panel, and Fig. 3, last panel). Our result is consistent with new HI 
data from the GALFA survey. 

Our HI absorption spectra clearly reveal reliable HI absorption 
features in the velocity range of 0 to 70$\pm$3 km s$^{-1}$ for W51C 
and 0 to 65$\pm$8 km $^{-1}$ for G49.2-0.35 but not beyond 73 km 
s$^{-1}$ for both sources. This reveals that W51C is at 
approximately the same distance as G49.2-0.35. The CO clump (A) at the velocity of 
71.2 km s$^{-1}$ sitting at the core of the HII region G49.2-0.35 
(Ceccarelli et al. 2011; also Carpenter \& Sanders 1998) is likely 
associated with the 65$\pm$8 km$^{-1}$ HI absorption feature towards 
G49.2-0.35. Thus it is possible that there exists an interaction 
between the W51C shock and the CO clump associated with HII G49.2-0.35. 
Because the HV HI gas is behind G49.1-0.38, which is far behind 
G49.2-0.35 and CO clump A, there is no interaction between W51C and 
the HV HI gas. If we still believe the HV HI gas is caused by 
shocks, it is more reasonable to believe that the shock is from 
stellar winds from the star forming region, i.e. HII region G49.1-0.38.       

\subsection{Are the OH masers from a SN Shock or from a star forming region?}
Two 1720 MHz OH masers are coincident with or adjacent to the
HII region G49.2-0.35 within W51B (see Fig. 1 of Brogan et al. 
2000). The velocities of the masers (69.2 and 72.1 km s$^{-1}$) are consistent 
with the RRL velocity (67.2$\pm$1 km s$^{-1}$) and the highest HI absorption 
velocity (65$\pm$8 km s$^{-1}$) of G49.2-0.35. So it is reasonable 
that the OH masers are associated with G49.2-0.35. The 
masers are located near the interface between SNR W51C and G49.2-0.35 
based on ROSAT X-ray observations of W51B and W51C, although the interface 
is not obvious (Koo \& Kim 1995). 
1720 MHz OH masers can be produced either in SNR-shocked clouds or 
in star forming regions (SFR) under similar conditions (Green et al. 
1997, Wardle 1999, Caswell 2004).  However, they have different pumping schemes: 
the SNR 1720 MHz OH maser seems to be the sole maser transition of OH with 
corresponding broad OH absorption at 1667, 1665 and 1612 MHz, while the 
SFR maser most commonly accompanies emission at the 1665 MHz OH transition. 
Hewitt et al.(2008, 2006) give examples: the 1720 MHz OH masers within the middle-aged 
/SNRs W28, W44 and IC443 are clearly associated with SNR shocks interacting with 
adjacent molecular clouds. 
But for W51C, there is 1665 MHz emission at 74 km/s, 
which indicates that the OH MHz masers likely originate from SFR activity. 
Many SFR 1720 MHz OH masers are accompanied by the 6035 MHz OH masers (Caswell 2004, Etoka et al. 2012), so additional 6035 MHz maser observations of W51C may help strengthen our conclusion.

Koo et al. (2010) made an HI Zeeman observation of the shocked 
HV HI gas of W51B and obtained a line-of-sight magnetic field of 
B$_{LOS}$ $\le$ 0.15 mG in the HV HI gas (above 85 km s$^{-1}$). 
Brogan et al. (2000) measured the B$_{LOS}$ of 1.5 $\sim$ 1.9 mG in 
the 1720 MHz OH maser gas overlapping W51B. This difference is 
easily understood if we believe the HI HV gas is not related with 
the OH maser gas, i.e. they are physically separated. This supports 
our above conclusion that the HV HI gas is behind the HII region 
G49.1-0.38. 

\subsection{Conclusion}
We have used HI absorption spectra to determine the distances to W51C (4.3 kpc) 
and the HII regions G49.2-0.35 (4.3 kpc) and G49.1-0.38(6.9 kpc). We also consider 
whether W51C is interacting with atomic gas. This is of interest because 
GeV and TeV emissions have been detected recently (Abdo et al. 2009) from the area
where W51B and W51C overlap. TeV emission from several middle-aged 
SNRs interacting with adjacent molecular clouds has been considered 
to originate from hadronic processes (e.g. Ohira et al. 2011).
However for W51C, we have found it not interacting with any HV HI.
Also we find that the observed nearby OH masers from dense molecular gas are 
likely associated with the star forming regions in W51B and not W51C. 
These new results show clear absence of evidence of molecular or atomic gas 
interacting with W51C, thus casting doubt on a hadronic origin for the GeV/TeV emission 
from W51C.  We think W51C is similar to the case of Tycho SN 1572 (Tian \& Leahy 2011),
where the SNR was found not to interact with molecular or HI clouds.  

\begin{acknowledgements}
WWT thanks the NSFC and China Ministry of Science and 
Technology under State Key Development Program for Basic Research 
for support. DAL acknowledge support from the 
Natural Sciences and Engineering Research Council of Canada. We 
thank Drs. Gibson and Koo for providing GALFA data. This 
publication was partly supported by a grant from the John Templeton 
Foundation and National Astronomical Observatories of the CAS. The 
opinions expressed in this publication are those of the authors do 
not necessarily reflect the views of the John Templeton Foundation 
of NAOCAS. The funds from John Templeton Foundation were awarded in 
a grant to The University of Chicago which also managed the program 
in conjunction with NAOCAS. Finally, we thank the referee for suggestions which
improved this paper.
\end{acknowledgements}

\end{document}